\documentclass[prl,twocolumn,superscriptaddress]{revtex4-1}
\usepackage{graphicx,color}
\usepackage{amsmath,amssymb,bm}

\usepackage[plainpages=false,pdfpagelabels,colorlinks=true,linkcolor=red,urlcolor=blue,citecolor=blue,pdftitle={Comment on ``Systematic Construction of Counterexamples to the Eigenstate Thermalization Hypothesis''},pdfauthor={},pdfdisplaydoctitle=true,pdfduplex=DuplexFlipLongEdge]{hyperref}

\begin{document}

\title{Comment on ``Systematic Construction of Counterexamples to the Eigenstate Thermalization Hypothesis''}
\author{Rubem Mondaini}
\affiliation{Beijing Computational Science Research Center, Beijing 100193, China} 
\author{Krishnanand Mallayya}
\affiliation{Department of Physics, Pennsylvania State University, University Park, Pennsylvania 16802, USA}
\author{Lea F. Santos}
\affiliation{Department of Physics, Yeshiva University, New York, New York 10016, USA}
\author{Marcos Rigol}
\affiliation{Department of Physics, Pennsylvania State University, University Park, Pennsylvania 16802, USA}

\maketitle

The Letter~\cite{shiraishi_mori_17} claims to provide a general method for constructing local Hamiltonians that do not fulfill the Eigenstate
Thermalization Hypothesis (ETH)~\cite{deutsch_91, srednicki_94, rigol_dunjko_08, rigol_srednicki_12, dalessio_kafri_16}. We argue that the claim is misguided.

The Letter~\cite{shiraishi_mori_17} reports the construction of block-diagonal Hamiltonians with nonlocal many-body conserved quantities. In the second example, one such quantity is used to construct a Hamiltonian with two exponentially large symmetry sectors (the on site magnetic fields and local spin interactions were chosen to be different in the two sectors). It is not surprising that the ETH is violated when mixing them. Random matrix theory, the base of our understanding of the ETH~\cite{dalessio_kafri_16}, only applies within each symmetry sector and not to the entire Hamiltonian~\cite{dalessio_kafri_16, santos_rigol_10a, *santos_rigol_10b, Gubin2012}. Consequently, the ETH should be studied within each sector separately.

\begin{figure}[!b]
\includegraphics[width=1\columnwidth]{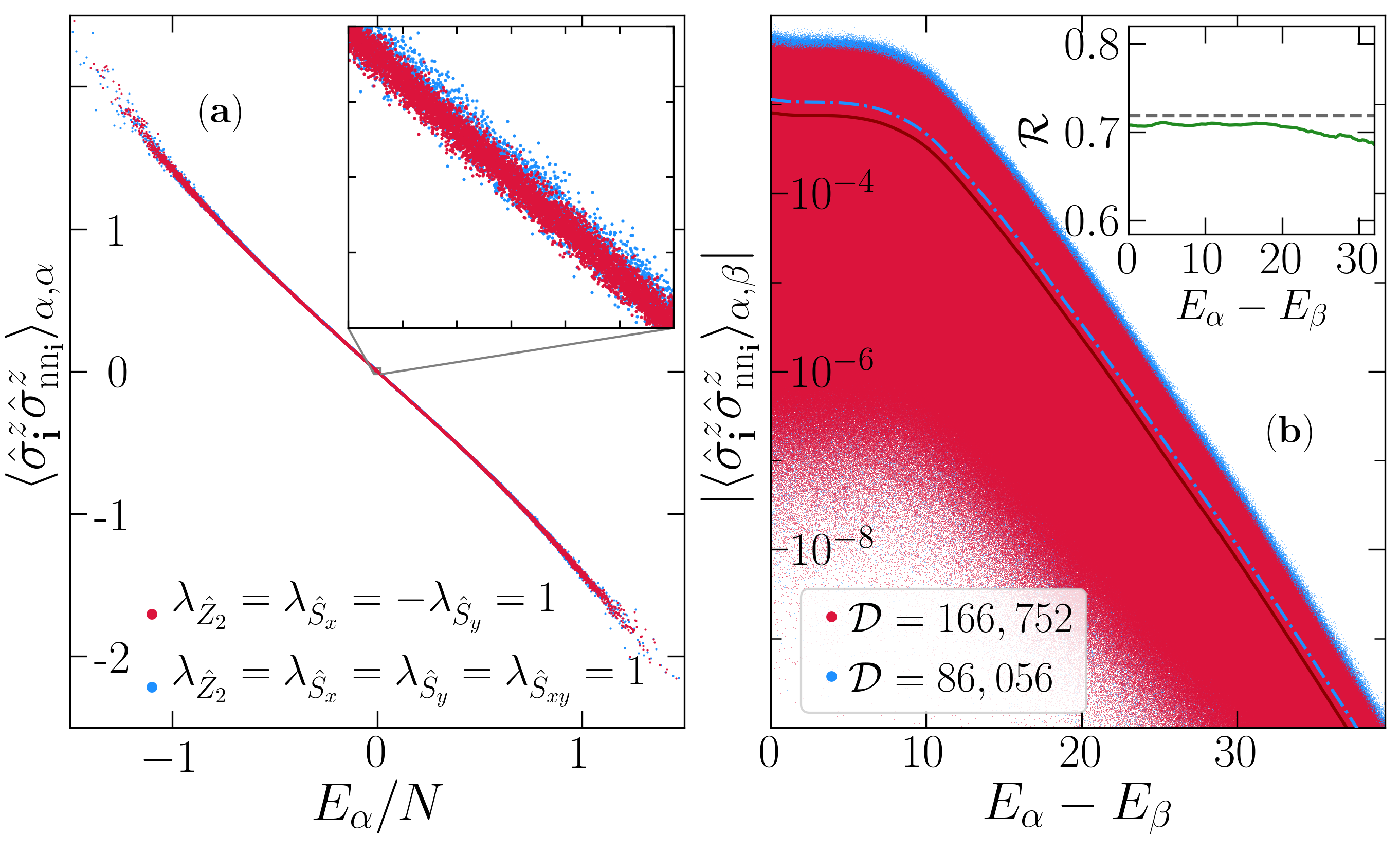}
\caption{(a) Diagonal and (b) off-diagonal (for $|E_\alpha+E_\beta|/N\leq0.1$) matrix elements of the nearest-neighbor spin correlations $\hat\sigma_{\bf i}^z\hat \sigma_{\text{nn}_{\bf i}}^z$ for the ferromagnetic transverse-field Ising model $(g=J)$ in two dimensions ~\cite{mondaini_fratus_16,mondaini_rigol_17}. Continuous lines in (b) depict running averages. The inset in (a) shows that the fluctuations of the diagonal matrix elements are different in the two sectors shown. Inset in (b): ratio $\cal R$ between the running averages of the off-diagonal matrix elements. The dashed line shows that, as expected from the ETH~\cite{dalessio_kafri_16, srednicki_99}, the ratio $\cal R$ is very close to the square root of the inverse ratio between the Hilbert space dimension ${\cal D}$ of the sectors. $\lambda_{\hat Z_2}$, $\lambda_{\hat S_x}$, $\lambda_{\hat S_y}$, $ \lambda_{\hat S_{xy}}$ are the eigenvalues of the spin-flip, mirror-$x$, mirror-$y$, and mirror along the $x=y$ line, symmetries, respectively. The results shown are for the zero momentum sector of a lattice with $N=5\times5$ sites~(see Ref.~\cite{mondaini_rigol_17} for further details).}
\label{fig}
\end{figure}

Conserved nonlocal many-body operators associated with lattice translations, point-group symmetries, and particle-hole transformations~\cite{santos_rigol_10a, *santos_rigol_10b, rigol_09a, *rigol_09b, steinigeweg_khodja_14, sorg_vidmar_14, kim_ikeda_14, beugeling_moessner_14, *beugeling_moessner_15, mondaini_fratus_16, mondaini_rigol_17} also generate block-diagonal Hamiltonians. In the chaotic regime of such models, in contrast to models with local conserved quantities (e.g., total particle number \cite{rigol_09a, *rigol_09b, steinigeweg_khodja_14, sorg_vidmar_14, kim_ikeda_14, beugeling_moessner_14, *beugeling_moessner_15}) and the models in Refs.~\cite{shiraishi_mori_17, mori_shiraishi_17, lan_powell_17, lan_vanhorssen_17}, the eigenstate expectation values of few-body operators are usually the same (up to finite-size effects) in different symmetry sectors [see Fig.~\ref{fig}(a)]. The need to analyze each sector separately becomes apparent when studying the off-diagonal matrix elements~\cite{mondaini_rigol_17}, an equally important part of the ETH~\cite{dalessio_kafri_16}. At any energy, the average magnitude of off-diagonal matrix elements of few-body operators that do not break symmetries of the Hamiltonian is generally different within different symmetry sectors. Also, they vanish between eigenstates that belong to different sectors. Hence, mixing different symmetry sectors may lead one to conclude that the ETH is violated while it is not.

In Fig.~\ref{fig}(b), we plot off-diagonal matrix elements, and their running average, within the same symmetry sectors as in Fig.~\ref{fig}(a). The mismatch of their magnitudes is apparent. Their ratio is determined by the ratio of the Hilbert space dimensions~\cite{dalessio_kafri_16,srednicki_99}, see inset in Fig.~\ref{fig}(b). This shows that different symmetry sectors should not be mixed when discussing the ETH.

We are also troubled by the statement in Ref.~\cite{shiraishi_mori_17} that numerical simulations have shown that the ETH is valid for Hamiltonians with: (i) translational invariance, (ii) no local conserved quantity, and (iii) local [$O(1)$ support] interactions. None of these conditions is necessary for the onset of quantum chaos and the validity of the ETH.

An early discussion on the connection between the ETH and thermalization in many-body lattice Hamiltonians involved a nontranslationally invariant system~\cite{rigol_dunjko_08}. Many of the models in which the ETH has been verified have a local conserved quantity, the total particle number or magnetization~\cite{rigol_09a, *rigol_09b, steinigeweg_khodja_14, sorg_vidmar_14, kim_ikeda_14, beugeling_moessner_14, *beugeling_moessner_15}. Finally, in Ref.~\cite{khatami_pupillo_13}, the ETH was verified in a model of hard-core bosons with dipolar ($1/r^3$) interactions in the presence of a harmonic trap, which does not satisfy any of the three conditions.

\begin{acknowledgments}
R.M. is supported by the NSFC, Grants No.~U1530401, No.~11674021 and No.~11650110441. K.M. and M.R. are supported by the NSF, Grant No.~PHY-1707482. L.F.S. is supported by the NSF Grant No.~DMR-1603418.
\end{acknowledgments}


\bibliography{references}

\end{document}